\documentclass{revtex4-1}
\usepackage{MnSymbol}
\usepackage{amsmath}
\usepackage{xparse}
\usepackage{tabularx}
\usepackage{multirow}
\usepackage{graphicx}

\newcommand{\anynumber}{\forall}
\def\P{\mathcal{P}}
\NewDocumentCommand{\p}{ O{} O{} m }{ {}_{#1} \P_{#3} #2 }
\NewDocumentCommand{\pA}{ O{} O{A} m }{\ensuremath{{}_{#1} \P_{#3}{\left\{#2\right\}} }}
\newcommand{\seroconvert}[1]{\mathcal{C}{\left\{#1\right\}}}
\newcommand{\ffrac}[1]{\seroconvert{A{+}#1}\frac{\pA[0][A{+}#1]{\op}}{\pA[0][A]{\op}}(V{+}\the\numexpr #1+1\relax T)}

\newcommand{\fhatterm}[0]{\seroconvert{A{+}i}\pA[0][A{+}i]{\op}\left(\nu{+}(i{+}1)\tau\right)}

\newcommand{\eref}[1]{Eq.~\ref{#1}}
\newcommand{\fref}[1]{Fig.~\ref{#1}}

\def\pH{\rho_H}
\def\tp{2^{{+}}}
\def\op{1^{{+}}}
\def\eg*{\textit{e.g.}}
\def\ie*{\textit{i.e.}}

\begin{document}
\title{Serostatus Testing \&\ Dengue Vaccine Cost-Benefit Thresholds}
\date{\today}
\author{Carl A. B. Pearson}
\email[direct correspondence to: ]{carl.pearson@lshtm.ac.uk}
\affiliation{Department of Infectious Disease Epidemiology \&\ Centre for Mathematical Modelling of Infectious Disease,\\ London School of Hygiene \&\ Tropical Medicine}
\affiliation{South African Centre for Epidemiological Modelling and Analysis,\\ Stellenbosch University}
\author{Samuel Clifford}
\author{Kaja M. Abbas}
\author{Stefan Flasche}
\affiliation{Department of Infectious Disease Epidemiology \&\ Centre for Mathematical Modelling of Infectious Disease,\\ London School of Hygiene \&\ Tropical Medicine}
\author{Thomas J. Hladish}
\affiliation{Department of Biology \&\ Emerging Pathogens Institute,\\ University of Florida}

\begin{abstract}
The World Health Organisation currently recommends pre-screening for past infection prior to administration of the only licensed dengue vaccine, CYD-TDV.  Using a bounding analysis, we show that despite additional testing costs, this approach can improve the economic viability of CYD-TDV: effective testing reduces unnecessary vaccination costs while increasing the health benefit for vaccine recipients.  When testing is cheap enough, those trends outweigh additional screening costs and make test-then-vaccinate strategies net-beneficial in many settings.

We derived these results using a general approach for determining price thresholds for testing and vaccination, as well as indicating optimal start and end ages of routine test-then-vaccinate programs.  This approach only requires age-specific seroprevalence and a cost estimate for second infections.  We demonstrate this approach across settings commonly used to evaluate CYD-TDV economics, and highlight implications of our simple model for more detailed studies.  We found trends showing test-then-vaccinate strategies are generally more beneficial starting at younger ages, and that in some settings multiple years of testing can be more beneficial than only testing once, despite increased investment in testing.
\end{abstract}

\maketitle

\section{Introduction}

Dengue virus inflicts substantial disease burden globally, particularly in lower-middle income countries \cite{bhatt2013,stanaway_2016}, and many organisations have advocated various control measures. However, since these settings have many competing development options to improve quality of life but limited resources, potential control efforts must be evaluated by return value.  One such dengue control tool, and the only currently licensed vaccine, CYD-TDV (commercially: Dengvaxia), presents a complicated assessment: there is potential value, but also safety risks and risk mitigation costs.

Dengue infections elicit a complex immune response, particularly in regions where people typically experience multiple infections over their lifetimes.  There are four known dengue serotypes; infection by one serotype confers apparently lifelong immunity to it, and temporary immunity to others. Disease risk varies substantially: first infections have moderate risk, second infections have relatively high risk, and post-second infections have low risk; this phenomenon is called antibody-dependent enhancement \cite{halstead2008dengue, wilder2019dengue}. To preclude enhancement, vaccine development has focused on creating a product effective against all serotypes. CYD-TDV initially appeared to satisfy this goal \cite{capeding_2014,villar_2015}, but later data suggested it acts more like an asymptomatic natural infection, providing transient immunity, but ultimately contributing only to immune history \cite{ferguson2016}. Presumably because second infections are more likely to cause disease, CYD-TDV appears to enhance disease risk in seronegative recipients (those having no prior dengue infections) while being highly efficacious for recipients with previous infections. These diverging outcomes present an ethics challenge for vaccine deployment.

A multi-model comparison study estimated that the broad use of CYD-TDV in high-burden settings (like those represented in the trials) would reduce the rate of both typical and severe cases\cite{flasche2016}. These findings contributed to the initial World Health Organisation (WHO) guidance: CYD-TDV is acceptable for use in settings with at least 50\% seroprevalence (population fraction with antibodies to at least one dengue serotype) in the target age for routine vaccination, and the vaccine should only be administered to recipients at least 9 years old \cite{who_2016}.

Additional observation in CYD-TDV trial populations confirmed increased risk of hospitalisation and severe dengue in seronegative vaccine recipients \cite{sridhar_2018}. The Strategic Advisory Group of Experts on Immunisation (SAGE) suggested this risk could be avoided by requiring evidence of past dengue infection based on serological testing \cite{WILDERSMITH2019e31}. In 2018, WHO revised their recommendations to follow this advice \cite{who_website_2018,worldhealthorganization_2018}. Practically, the revised guidance entails a point-of-care rapid diagnostic test (RDT); the lack of such a test (as of March 2019) and concerns over price have stalled deployment of the recommended test-then-vaccinate strategy.

At the generally assumed price point of 78 USD, CYD-TDV is a marginal investment \cite{flasche2016}, so adding testing costs would seemingly decrease its attractiveness further. But screening can also enable more optimal vaccine use, potentially preventing more high risk infections and avoiding vaccination when unlikely to be beneficial.  To inform ongoing development of RDT kits and CYD-TDV deployment plans, we investigated this possibility using limiting assumptions. We found pre-screening strategies can provide net benefit compared to both no vaccination and vaccination without testing, but generally have highest returns when started earlier than the recommended 9-year-old threshold. We also often found greater benefits when initially seronegative candidates are periodically re-tested for seroconversion. Our results provide useful cost bounds for tests, and indicate promising areas for further analysis, such as the appropriate timing of repeated testing.

\section{Methods}

We derive cost thresholds using two limiting assumptions. We ignore dengue transmission, and thus any indirect vaccine effect, and assume the test provides perfect information. Treating dengue more like an environmental factor is justified given the setting (\ie*, endemic with high force infection) and prior work estimating limited impact on transmission from CYD-TDV \cite{flasche2016}. The assumption for test performance, while insufficient for detailed cost estimates, can provide a maximum \emph{cost threshold}; real tests will misclassify individuals, thus under-performing on benefit, and so must be within the limiting price to have any prospect of returns.

\subsection{Dengue Disease \&~CYD-TDV Models}

\begin{figure}
\centering
\includegraphics[width=\linewidth]{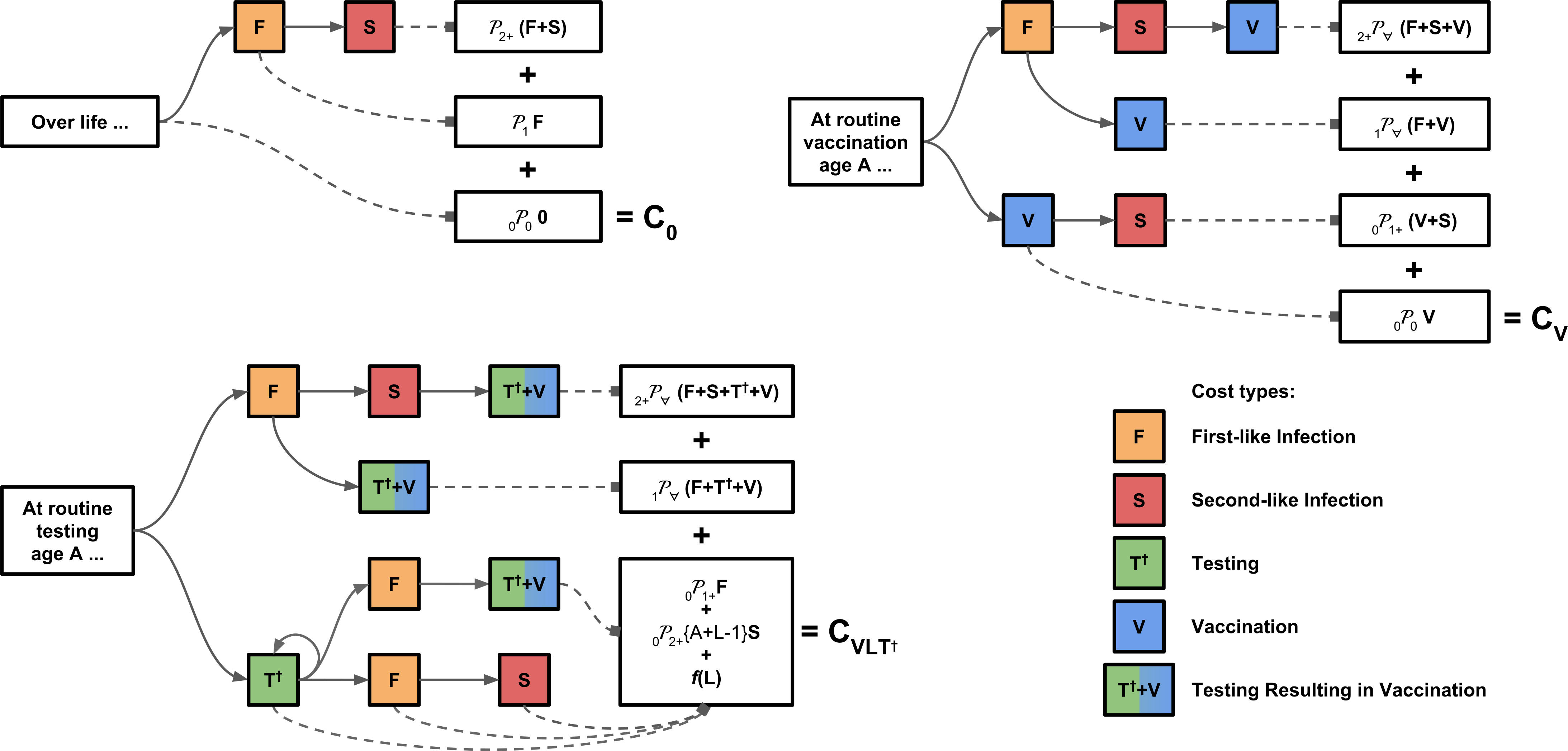}
\caption{\textbf{Lifetime Outcomes By Intervention}. Trajectories for no vaccination versus vaccination without testing versus vaccination and multiple testing. Each path represents a possible life history, resulting in ultimate outcomes weighted by the probability of following that path. For detailed breakdown of the branching probabilities, see the additional annotations on Supplement diagrams Fig.~S2-4 and S7-8.\label{fig:diagram}}
\end{figure}

We represent dengue disease as having three infection outcomes: first-like, second-like, and post-second-like. Without vaccination, these correspond to the average burden of first, second, and subsequent infections, respectively. Note this average includes asymptomatic infections, not just observed cases that present clinically. We assume that first-like infections are lowest cost, second-like are highest, and post-second-like have no cost, reflecting their respective disease severity. We assume CYD-TDV replaces one of these outcomes; thus, ideal testing strategies enable replacing additional second-like outcomes. \fref{fig:diagram} shows the possible life trajectories without vaccination, with unconditional vaccination, and with vaccination of only seropositives.

\subsection{Vaccination, Testing, and Cost Model}

We calculate costs per individual, using the average costs of first and second infections ($F$, $S$) and probabilities of various life trajectories. We represent those probabilities by $\pA[N]{X}$, meaning the probability of $N$ infections by vaccination age $A$ and $X$ total lifetime infections. We also define $\seroconvert{A}$, the probability of seroconverting between age $A$ and $A+1$, conditional on seronegativity at $A$. We estimate these probabilities with the exposure model described in the next section; the costs are also context-specific. We represent the costs of vaccination ($V$) and testing ($T$) with a \emph{binary} mechanism, which detects any past infection.  We also considered an \emph{ordinal} mechanism, which detects the number of past infections; see the Supplement for more details. Finally, we allow for up to $L$ tests, one each year from age of first consideration for vaccination.

To provide net benefit, a binary test must satisfy the cost constraint in \eref{eqn:del_VLTdag}; the Supplement has the full derivation, as well as results for ordinal tests.  For any intervention with testing in our model, we find that first infection cost, $F$, does not impact these constraints.  Therefore, we are able to generalize our framework across settings by expressing intervention costs as a fraction of secondary infection costs: $\nu=\frac{V}{S}$ and $\tau=\frac{T}{S}$.

\begin{align}
\hat{f}(L) = \pA[0][A{+}L{-}1]{\anynumber}(L{-}1)\tau &+ \sum_{i=0}^{L-2}\fhatterm{}\\
\tau + \p[\op]{\anynumber}\nu + \hat{f}(L) &\leq \left(\p{\tp}-\p[\tp]{\anynumber}-\pA[0][A+L-1]{\tp}\right)\label{eqn:del_VLTdag}
\end{align}

All intervention costs are on the left of \eref{eqn:del_VLTdag}; the net benefit is thus the right side minus the left. For results, we report return on investment (ROI), \ie* net benefit over intervention cost:

\begin{align}
\frac{\p{\tp}-\p[\tp]{\anynumber}-\pA[0][A+L-1]{\tp}}{\tau + \p[\op]{\anynumber}\nu + \hat{f}(L)} - 1 \geq 0,\quad\textrm{for positive returns}\label{eqn:ROI}
\end{align}

To identify circumstances where adding testing increases intervention benefits, we compared vaccination with and without testing, which produces a similar equation, now also dependent on $F$:

\begin{equation}
\tau + \p[\op]{\anynumber}\nu + \hat{f}(L) \leq \left(\p[0]{\op}-\pA[0][A+L-1]{\tp}\right) -\p[0]{\op}\frac{F}{S} + \nu
\end{equation}

\subsection{Dengue Infection \& Intervention Model}

We represent dengue exposure in annual increments: each year, an individual is potentially exposed. We divide the population into risk groups, low and high. Thus, the exposure model has three parameters: $p_H$, population fraction at high risk; and $f_X = 1 - s_X$, the exposure (or complementary survival) probability per year for low ($X=L$) and high ($X=H$) risk populations, which means the probability of being seropositive at age $A$ is:

\begin{equation}
\mathbf{P}_{{+}}\left\{A\right\} = \pH\left(1-s_H^A\right) + \left(1-\pH\right)\left(1-s_L^A\right)
\end{equation}

This model could be fit to age-seroprevalence data with several different techniques; here we used a maximum likelihood based approach (see Supplement for details).  We fit to both yearly \cite{morrison2010} and aggregated \cite{lazou2016} age cohorts of seropositive count and sample size.  Because the model has three parameters, there must be three or more points to constrain the fit.

With exposure parameters, we can simulate infections to estimate life trajectory probabilities. We assume four dengue serotypes; each year we uniformly select one for exposures, then probabilistically expose individuals based on their risk category. Exposure leads to infection, unless the individual (1) has previously been infected by this serotype or (2) was infected in the previous year.

We used data from CYD14 \cite{lazou2016} and long term study of Iquitos, Peru \cite{morrison2010} to develop two practical examples for Malaysia and Peru, and to more generally bound the region of relevant \emph{force of exposure} (represented by seroprevalence in 9-year-olds) and \emph{disparity} (combination of high risk fraction and the odds seropositivity in 9-year-olds).

\section{Results}

\begin{figure}
\centering
\includegraphics[width=\linewidth]{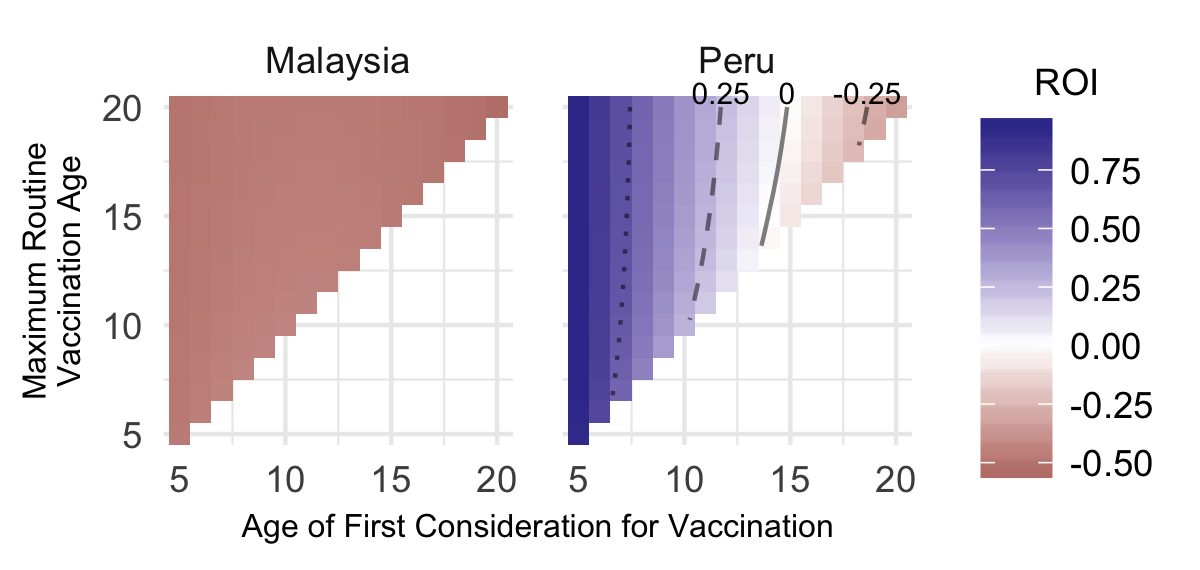}
\caption{\textbf{Vaccination \& Testing versus Status Quo}. Two settings with seroprevalence between 70 and 80\%. Comparing these two settings using the assumed vaccination and testing costs, we see that in settings where second infection cost $S$ is low and infection disparity is high (as assumed for Malaysia), ROI is negative. However, in settings where $S$ is high and infection disparity is low (as assumed for Peru), ROI is positive over a broad range, when vaccination starts early enough.\label{fig:1}}
\end{figure}

Previous work assumed vaccination (all doses plus logistics) costs 78 USD \cite{flasche2016}, and initial test cost estimates are 5 USD. Using the same South East Asian and Latin American societal costs as in previous work, we have for Malaysia $\left\{\tau,\nu\right\}=\left\{0.06,0.91\right\}$ and for Peru $\left\{0.02,0.35\right\}$. Using the transmission parameters derived for these regions, we obtained ROI surfaces for initial and maximum vaccination age (\fref{fig:1}). These two settings highlight very different potential benefits surfaces, as well as the importance of considering age of vaccination when evaluating the strategies.

\begin{figure}
\centering
\includegraphics[width=\linewidth]{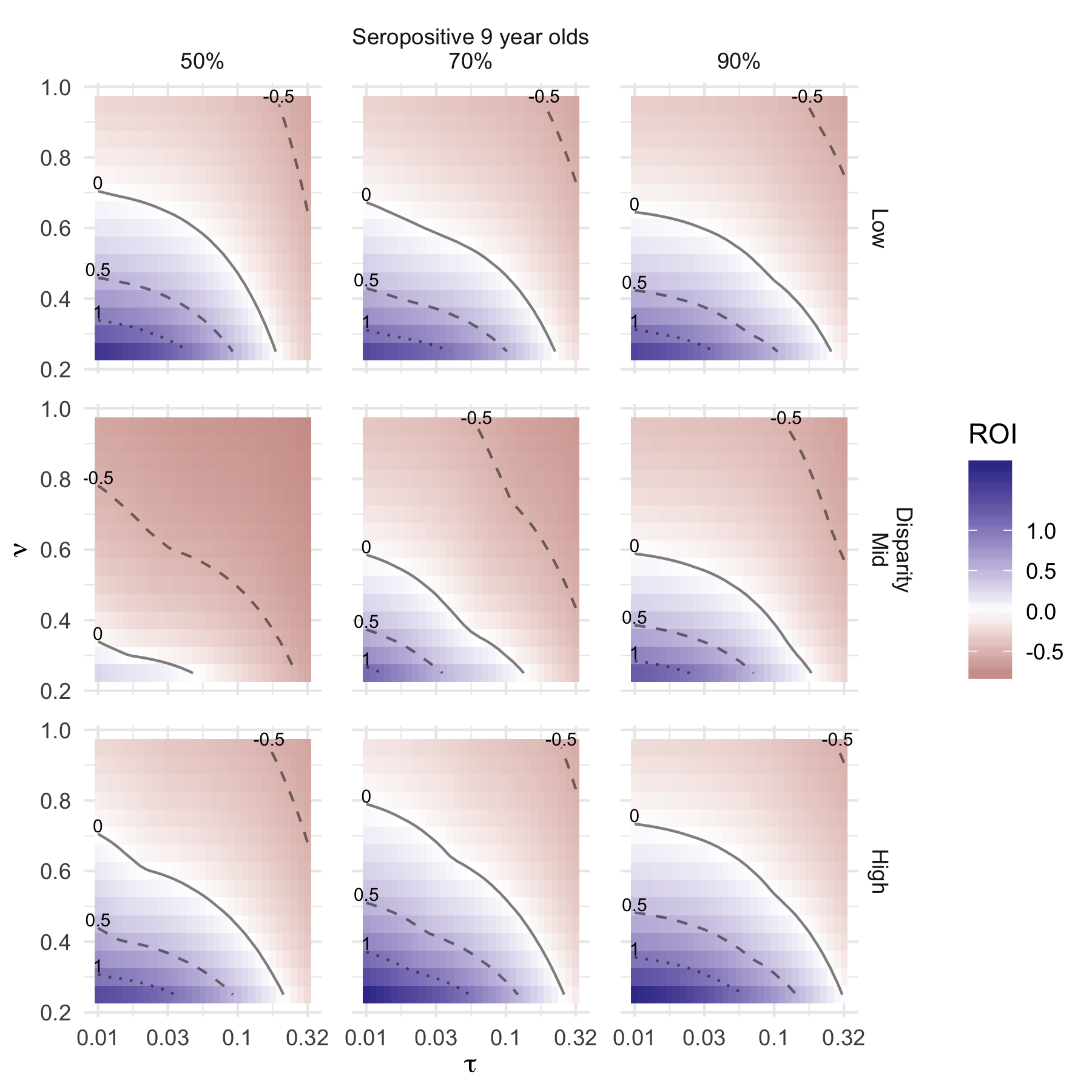}
\caption{\textbf{General Survey of Settings \& Costs}. Cost-benefit surfaces for a range of settings. In general, more benefit can occur in higher transmission settings. The disparity response is more complicated; disparity is a mix of how small the high risk population is and how much additional exposure they receive. By the mid setting, the high risk population is close to maximum annual exposure rate, so shrinking that population further in the high disparity means the low risk exposure rate must go up to maintain the target seropositive rate.\label{fig:2}}
\end{figure}

\begin{figure}
\centering
\includegraphics[width=0.5\linewidth]{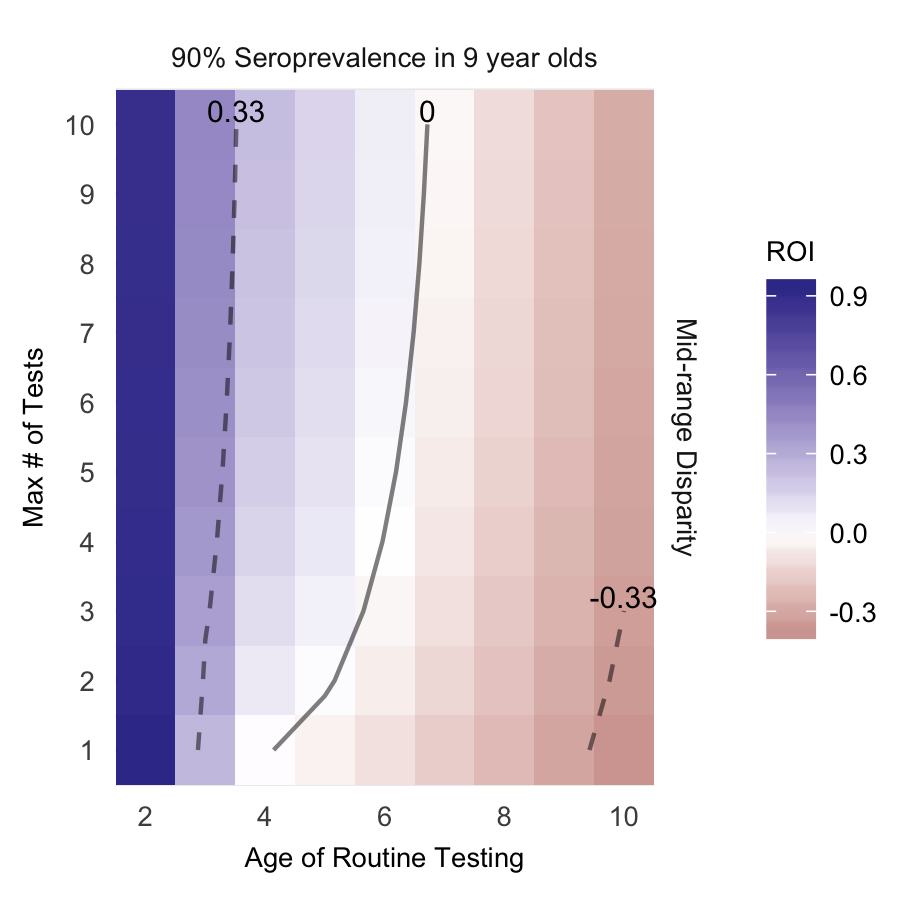}
\caption{\textbf{Sensitivity to Number of Tests}. Shows an area of epidemiological and economic parameter space ($\tau=0.01$, $\nu=0.5$) where getting positive or negative returns depends on number of tests allowed by the strategy.\label{fig:testresults}}
\end{figure}

We also calculated general sensitivity to context; in \fref{fig:2} we show the best possible age and testing duration combination for the parameters (for vaccination of 5- to 20-year-olds, testing at most 10 years).  When we look more closely at particular settings to evaluate general trends in routine testing age and maximum number of tests, there are some nuances.  \fref{fig:testresults} highlights the common trend that testing earlier is better, but this effect is sensitive to detailed context.  This particular region highlights a {\em minimum} number of tests to achieve net benefits; in other areas, there is a maximum.  We found these kind of trend reversions across several parameters.

\section{Discussion}

Several promising new dengue vaccines are currently in trials, but CYD-TDV was once similarly promising.  These alternatives may ultimately present better solutions, but rather than waiting on the promise of the next product, we should determine how to use CYD-TDV now.  Given the complex dengue immunology and vaccine development history, the best assumption is that these new vaccines will also have quirks.  However, there may still be utility in flawed vaccines.  We have developed a convenient approach to identify settings and deployment scenarios worthy of more detailed assessment, while clearly precluding some settings or test prices from further consideration.

Our modelling work provides two key insights about using CYD-TDV with RDTs, and potentially other in-development vaccines.  First, multiple testing can improve a program's cost effectiveness, and may even be required to achieve net benefit, depending on test price and realized test performance.  Multiple testing will impose some additional costs (\eg* records keeping), but may enable economies of scale.  Though we assumed a perfect test, a real product will sometimes misclassify potential vaccinees.  The net effects of logistics and test error for multiple testing are not obvious, but the potential merits of multi-test strategies warrants more detailed future work.

Second, given multiple testing, routine test-then-vaccinate programs provide better ROI the younger the age they target for many relevant settings, including below the current recommended threshold of 9 years.  This will also be sensitive to test performance, and particularly how that performance depends on target population seroprevalence, but the current age threshold for vaccination should be revisited.  Despite current licensing, CYD-TDV trials included younger participants and, after controlling for seropositivity, seem to have shown safe and efficacious outcomes.

These insights should inform further work about test-then-vaccinate strategies. More detailed models could address indirect effects (positive and negative) of expanding vaccine coverage, as well as any age-specific effects for younger vaccination. Using estimated test performance will also shrink the positive ROI region; understanding the magnitude of this effect will be critical to final deployment decisions. Though we conclude multiple testing has superior performance in many settings, we did not address the appropriate interval for testing. In our model world, it is impossible to get two infections in a year (the assumed interval for testing), thus testing every other year would be more effective. Varying this periodicity could expand the positive ROI region, though incorporating test performance will complicate this effect.

In summary, though CYD-TDV may not be the hoped-for durable, efficacious, tetravalent dengue vaccine, with the addition of testing it could provide a cost-beneficial intervention and satisfy medical ethics requirements.  In settings where our constraint model indicates testing and vaccinating may be beneficial, more detailed study is warranted to estimate precise ROIs by relaxing our simplifying assumptions.

\section{Acknowledgements}

CABP is supported by EBOVAC3. Contributions by SC and SF are supported by Wellcome Trust. KMA is supported by NIH/NIGMS R01 GM109718. TJH is supported by NIH/NIGMS U54 GM111274.

\bibliography{refs}

\end{document}